\documentclass[11pt]{article}  
\usepackage{graphicx,floatflt,amssymb}    
\usepackage{epsfig}    
\usepackage{graphics}    
\usepackage{psfrag}    
\newcommand{\ba}{\begin{array}}     
\newcommand{\ea}{\end{array}}     
\newcommand{\bd}{\begin{displaymath}}     
\newcommand{\ed}{\end{displaymath}}     
\newcommand{\be}{\begin{equation}}     
\newcommand{\ee}{\end{equation}}     
\newcommand{\bea}{\begin{eqnarray}}     
\newcommand{\eea}{\end{eqnarray}}

\begin{document}     
\vspace*{-0.5in}     
\renewcommand{\thefootnote}{\fnsymbol{footnote}}     
\begin{flushright}     
LPT Orsay/05-03 \\     
SINP/TNP/05-23\\ 
DNI-UAN/05-91FT\\    
\texttt{hep-ph/0511275}      
\end{flushright}     

\vskip 5pt     

\begin{center}     
{\Large {\bf Neutrinos in the simplest little Higgs scenario and TeV 
leptogenesis}}
\vskip 25pt     
{\bf Asmaa Abada$^{1}$},     
%\footnote{E-mail address: abada@lyre.th.u-psud.fr}      
{\bf Gautam Bhattacharyya$^{2}$},       
%\footnote{E-mail address: abada@lyre.th.u-psud.fr} 
%{\rm and} 
 {\bf Marta Losada$^{3}$}    
%\footnote{E-mail address: gb@theory.saha.ernet.in}       
\vskip 10pt      
{\small  
$^{1)}${\it LPT,      
Universit\'e de Paris-Sud XI, B\^atiment 210, 91405 Orsay Cedex,     
France} \\
$^{2)}${\it Saha Institute of Nuclear Physics, 1/AF Bidhan Nagar,      
Kolkata 700064, India}} \\     
$^{3)}${\it Centro de Investigaciones, 
Universidad Antonio Nari\~{n}o, \\ Cll. 58A No. 37-94, Santa Fe de Bogot\'{a},
Colombia}

\normalsize     
\end{center}

\vskip 20pt     

\begin{abstract}  
  The little Higgs scenario may provide an interesting framework to
  accommodate TeV scale leptogenesis because a TeV Majorana mass of
  the right-handed neutrino that we employ for the latter may find a
  natural place near the ultraviolet cutoff of the former. In this
  work we study how a light neutrino spectrum, generated radiatively,
  and TeV scale leptogenesis can be embedded in the simplest little
  Higgs framework. Alternatively, we highlight how the neutrino Yukawa 
  textures of the latter are constrained.

\vskip 5pt \noindent  
\texttt{PACS Nos: 14.60.St, 11.30.Fs, 14.80.Cp} \\  
\texttt{Key Words: Neutrino, Leptogenesis, Little Higgs}  
\end{abstract}

\renewcommand{\thesection}{\Roman{section}}  
\setcounter{footnote}{0}  
\renewcommand{\thefootnote}{\arabic{footnote}}

\section{Introduction}     
The Standard model (SM) with right-handed (RH) neutrinos provides an elegant
mechanism for thermal leptogenesis.  These RH neutrinos may also be
instrumental in generating masses and mixings for the light neutrinos through
the see-saw mechanism. There are two intertwined requirements: first,
reproduce the spectrum for light neutrinos in the observed range, and second,
generate enough CP asymmetry through the out-of-equilibrium decay of heavy RH
neutrinos (\cite{FY}-\cite{di}). This asymmetry can be transmitted to the
baryonic sector through sphaleron induced processes to explain the baryon
asymmetry of the Universe.  To achieve these two requirements, the RH
neutrinos should have Majorana masses and they should couple to the
left-handed (LH) lepton doublets and the SM Higgs via complex Yukawa
couplings. For natural choices of such couplings, in theories with a large
cut-off scale, the RH Majorana masses turn out to be quite close to the GUT
scale ($\sim 10^{16}$ GeV).  An alternative and attractive mechanism would be
to consider RH neutrinos at the TeV scale (\cite{pilaftsis2}-\cite{al}).  This
scale is accessible in ongoing and near future colliders. Moreover,
interesting new physics (like supersymmetry, extra dimensions, etc) could be
revealed around that scale.  It is now known that with three RH neutrinos it
is difficult to achieve TeV scale leptogenesis and reproduce at the same time
the small LH neutrino masses \cite{al}, unless one considers quasi-degenerate
Majorana neutrinos where the exact degeneracy is lifted by
e.g.~renormalisation group effects \cite{pilaftsis2} or small fine-tuning
\cite{Hambye2}.  In the present analysis, we first introduce a novel way of
generating a lepton asymmetry with 3 RH neutrinos at the TeV scale and
secondly, we adapt the scenario proposed in \cite{al}, which is a simple
extension of the Fukugita-Yanagida model \cite{FY} by introducing a fourth
generation in addition to the existing three of the SM plus a RH neutrino for
each of the four families.

Little Higgs models, in which the SM Higgs doublet is conceived as a
pseudo-Goldstone boson of a larger symmetry group, may provide an interesting
framework to accommodate TeV scale leptogenesis because the UV cutoff of such
models is also around a few TeV.  In this paper we study how a light neutrino
spectrum and TeV scale leptogenesis with RH neutrinos can be embedded in the
simplest little Higgs framework \cite{Schmaltz1,Schmaltz2}.

\section{Neutrinos in the simplest little Higgs scenario}
In the simplest little Higgs scenario the SM gauge group is enlarged to
 ${\rm SU(3)}_W\times {\rm U(1)}_X$ which entails the fermion doublets to be
 extended to triplets. The left-handed SU(3) lepton triplet is
 expressed as
  \bea \psi_{iL}=\left( \nu_i, \ell_i, N_i\right)_L^T ,
\eea where $i$ corresponds to a generation index. The minimal choice for the 
RH SU(3)
 singlet components are: 
\bea 
n_{iR} \quad{\rm and} \quad \ell_{iR}.
\eea 
Two SU(3) scalar triplets $\Phi_1$ and $\Phi_2$ are employed for the
spontaneous breaking of the ${\rm SU(3)}_W$ gauge symmetry to ${\rm SU(2)}_W$
with the vacuum expectation values (vevs) $f_{1,2}$ around the TeV scale. Now,
$\Phi_{1,2}$ can be expressed as
\bea \Phi_{1,2}=\rm{exp}\bigl(\begin{array}{c}\pm i
{\Theta\over f_{1,2}} \end{array}\bigr)\left(\begin{array}{c}0\\0\\
f_{1,2}+\rho_{1,2}\end{array}\right)\ ,
\label{phi12} 
 \eea
where $\rho_{1,2}$ are radial excitations on which we comment later. The
phase $\Theta$ is given by (keeping only the SM Higgs field components)
\bea 
\Theta={1\over
\sqrt{2}}\left(\begin{array}{c c c}0 &0& h^+ \\0 &0& h^0\\ h^- &h^{0*}&
0 \end{array}\right) , 
\label{theta}
\eea 
which contains the SM Higgs doublet $h=(h^+\ h^0)^T/\sqrt{2}$ \ .  Given the
aim of our work, we take $f_1=f_2=f$ for simplicity. The little Higgs
framework requires $f \sim 4\pi v$, where $v$ is the vev of the SM Higgs
doublet.

We arrange that only $\Phi_1$ couples to the lepton triplet through the
Yukawa interaction (similar to the approximation used in \cite{delaguila})
\bea
-{\cal{L}}_{Y} =  n_{iR}^c \lambda_{ij}{\psi}_{_{j
 L}} \tilde{\Phi}_1^\dagger +
{\rm h.c.}\ ,
\label{lag-yuk} 
\eea
where $\tilde{\Phi}_1$ is obtained by replacing the Higgs doublet by
$\tilde{h} \equiv i\tau_2 h^*$ in Eq.~(\ref{theta}), and $(i,j)$ run over
families. Since this Lagrangian involves only one scalar triplet, the global
axial SU(3) remains unbroken, and hence there is no contribution to the Higgs
mass divergence from this Yukawa sector.  Expanding the above Lagrangian in
terms of the SU(2) doublets $L_i=\left( \nu_i, \ell_i\right)_L^T$ and the
SU(2) singlets $N_i$, up to the second order in the field $h$, yields (after a
slight redefinition, $\psi_L^T \equiv (-iL,N)$)
\bea
 -L_{Y} = - \bar n_{iR}\lambda_{ij}{L}_{jL} \tilde{h}^\dag 
 + f \left(1 - \frac{h^\dagger h}{2f^2}\right)\bar n_{iR}
\lambda_{ij}N_{jL} + {\rm h.c.}\ .
\label{su2-lag} 
\eea

\section{Neutrino mass matrix and eigenvalues}
Taking into account the neutrino fields of the model we build the mass matrix.
There are two LH states $\nu_L$, $N_L$ and one RH state $\nu_R$ for each
generation.  We assume that the gauge singlet field $n_R (\equiv n^c_L)$ has a
Majorana mass $M$ around the TeV scale.  The mass matrix (tree level) in the
basis $\left(\nu_L, N_L, \nu^c_L \right)$ turns out to be
\bea
{\cal{M}}=\left(\begin{array}{c c c}0 &0&m_D\\
 0 &0&M_D\\m_D^\dag & M_D^\dag &M
\end{array}\right) , 
\label{m1}
\eea
where, $m_D= -\lambda v$ and $M_D=\lambda f\left(1-v^2/2f^2\right)$.  Strictly
speaking, each entry in the mass matrix should be interpreted as a matrix over
the generation indices. But for the ease of illustration we concentrate on a
single family.  The above mass matrix yields two massive and one massless
eigenstates with eigenvalues $\sim$ $M$, $M_D^2/M$, $0$. The eigenstate which
is dominantly the SM neutrino ($\nu'= \nu + (v/f) N$) is massless at this
stage.  But one-loop radiative corrections, obtained by integrating out the RH
singlet neutrino field, generates a dimension-5 effective operator (valid up
to the scale $M$)
\bea
L_5 \sim \frac{\lambda^2}{16\pi^2} \frac{(\Phi_2^\dag
\psi_L)(\Phi_2^\dag \psi_L)}{M},
\label{l5} 
\eea
which is realised through Yukawa interaction in conjunction with the scalar
quartic interaction $|\Phi_1^\dag \Phi_2|^2$ (the latter interaction is
generated at one-loop). It is straightforward to see by expanding the triplet
fields in the above operator in terms of the fields under SU(2) representation
that the zeros of the first two by two block of the mass matrix (\ref{m1}) are
now lifted.  The modified mass matrix now reads ($c \equiv 1/16\pi^2$; the
loop factor containing the Higgs quartic coupling $\lambda_h$ is order one and
we do not display it for simplicity)
 \bea
 {\cal{M}}=\left(\begin{array}{c c c} 
 c\cdot m_D^2/M & - c\cdot m_D M_D/M & m_D\\
 - c\cdot m_D M_D/M & c\cdot M_D^2/M & M_D
\\ m_D^\dag & M_D^\dag & M
\end{array}\right) . 
\label{m2} 
\eea

The mass matrix (\ref{m2}) has the following eigenvalues:
\bea
\label{evs}
 M_1 \simeq cm_D^2/2M; ~~ M_2 \simeq  M_D^2/M; ~~M_3 \simeq M. 
 \label{eigenvalues}
\eea 
It is not difficult to see why the modified eigenvalues are as in
Eq.~(\ref{evs}). Radiative corrections, which come with the factor `$c$' in
the first two by two block of Eq.~(\ref{m2}), are small enough to appreciably
alter the previously obtained nonzero eigenvalues $M$ and $M_D^2/M$ obtained
from diagonalisation of Eq.~(\ref{m1}).  Clearly, the nonzero determinant of
the matrix (\ref{m2}) immediately points to the smallest eigenvalue $M_1$ in
Eq.~(\ref{evs}).  It is straightforward to see that $M_1$ corresponds to the
light SM neutrino mass, $M_2$ is the Majorana mass of the state which is
dominantly $N_L$, while the RH singlet weighs around $M_3 \sim$ TeV.  Note
that even though $N_L$ is an SU(2) singlet, it acquires a Majorana see-saw
type mass. This happens because $N_L$ is a component of the SU(3) triplet
which experiences the interaction in Eq.~(\ref{l5}).  It is to be noted that
the light active neutrino masses have been generated through radiative
corrections and not by the conventional see-saw.

 Now we see how light active neutrino data constrain the different parameters.
 We take $M \in [0.5 - 1]$ TeV. This range is motivated by the requirements of
 TeV scale leptogenesis, as we shall see later.  A light neutrino eigenvalue
 $\sim \sqrt{\Delta m^2_{\rm atm}} \simeq 0.05$ eV implies the Yukawa coupling
 $\lambda \sim 10^{-5}$. The Majorana (Dirac) mass of the SU(2) singlet field
 turns out to be in the keV (MeV) scale.  At this point, it is worth comparing
 our analysis with that of \cite{delaguila}. The aim of \cite{delaguila} was
 to generate light neutrino masses with {\em unsuppressed} Yukawa couplings,
 which requires the {\em RH Majorana masses in the keV range}. On the
 contrary, we assume that lepton number is violated at the TeV scale. As a
 result, our Yukawa couplings for the light neutrinos are suppressed. As we
 shall see later, we can achieve a successful leptogenesis through the decay
 of RH TeV scale Majorana neutrinos.

\section{TeV scale leptogenesis}
Since the intrinsic scale of the little Higgs scenario is around a TeV, one
may ask whether all the requirements for a successful TeV scale leptogenesis
can be accommodated within such a framework. First, we recall that there are
two approaches in the literature to realise TeV scale leptogenesis: (a)
consider a quasi-degenerate spectrum of heavy RH neutrinos and enhance CP
asymmetry through resonant effects \cite{pilaftsis}; (b) extend the phase
space parameters, either (i) by admiting, for example, extra couplings that
allow three body decays of the RH neutrinos leading to an enhancement of CP
asymmetry \cite{hambye3}, or, (ii) by extending the particle content. As
regards the latter possibility, one may adopt among others either of two
approaches: (1) consider a supersymmetric framework \cite{Giudice}, (2)
minimally extend the SM by having a fourth chiral generation and add a heavy
RH neutrino for each of the four generations, assuming that the lepton
asymmetry is due to the decay of the lightest RH neutrino ($n_{R_1}$) in the
TeV scale \cite{al}.

 There are two key points: (a) the CP
 asymmetry
\bea 
\epsilon_1 = \frac{1}{8\pi [\lambda \lambda^\dag]_{11}}
 \sum_{j\neq 1} {\mathrm Im}{ [\lambda \lambda^\dag]_{1j}^2}
 f(M_{n_{R_j}}^2/M_{n_{R_1}}^{2}),
\label{eps1}
\eea
 where $f$ is the loop factor \cite{roulet}, given by  
\bea f(x) &=& \sqrt x \left[1 - (1+x)\ln\frac{1+x}{x} + \frac{1}{1-x}
\right], 
\label{floop}
\eea  
has to be magnified by the presence of a large Yukawa coupling, and (b) the
condition of out-of-equilibrium decay of ${n_{R_1}}$ has to be ensured, i.e.,
\bea
 \Gamma_{n_{R_1}}=\frac{(\lambda \lambda^\dag)_{11} 
{M_{n_{R_1}}}}{8\pi} <
 {H(T=M_{n_{R_1}})},
\label{gamma} 
\eea
where $H$ is the Hubble expansion rate at $T=M_{n_{R_1}}$.  These conditions
restrict the size of the Yukawa couplings.  In the following, we discuss two
possible scenarios for TeV scale leptogenesis in the context of the simplest
little Higgs model.

\subsection{Scenario $\mathbf I$: A 3-generation case}
First we consider the case with only 3 generations of fermions. We emphasize
that this is a novel approach to realize the marriage between TeV scale
leptogenesis and little Higgs. We take the first two generation of RH
neutrino masses $M_{n_{R_1}}< M_{n_{R_2}}$ of the order of a TeV, thus the
associated light LH neutrino masses arise through (\ref{m2}). The
corresponding Yukawa couplings are therefore suppressed $\sim 10^{-5}$.  The
third generation of RH neutrino in contrast has a much smaller Majorana mass
on the order of $M_{n_{R_3}} \sim$ keV, and the associated light LH active
neutrino acquires mass via the mechanism of Ref.~\cite{delaguila}. This
mechanism relies on the same operator of Eq.~(\ref{l5}), but with lepton
number violating mass $M\sim$ keV and {\em importantly with an unsuppressed}
Yukawa coupling $\lambda_{33} \sim 1$. We note here that $\lambda_{31}$ and
$\lambda_{32}$ can also be order one, or a little smaller, e.g. order 0.1 to
ensure the validity of the mechanism of \cite{delaguila}.

We consider the decay of the (next to lightest) RH neutrino ($n_{R_1}$) which
weighs around a TeV.  All Yukawa couplings entering the decay rate, i.e.
$\lambda_{1i}$, for all $i =1,2,3$, have to be order $10^{-7}$ or smaller to
ensure an out-of-equilibrium decay, see Eq.~(\ref{gamma}). Thus the
constraints on $\lambda_{1i}$ are about two orders of magnitude stronger than
those obtained from light active neutrino masses.  Thus the Yukawa matrix
(rows corresponding to the three RH neutrinos, columns to the active LH
neutrinos) looks like
 \bea
 {\cal{\lambda}}=\left(\begin{array}{c c c} 
 \epsilon  & \epsilon & \epsilon \\
 \beta & \beta & \beta \\
 \alpha & \alpha & {\cal{O}}(1) \\
\end{array}\right) , 
\label{lam3} 
\eea
where $\epsilon \sim 10^{-7}$, $\beta \sim 10^{-5}$, and ${\cal{O}}(0.1)
<\alpha< {\cal{O}} (1)$. It is understood that all entries are subject to
order one uncertainties. Clearly, all entries of the $(3 \times 3)$ light
active neutrino masses, proportional to $\lambda^T \lambda$ assuming the RH
mass matrix to be diagonal, receive contributions from both the mechanisms
cited above.

Now we consider the CP asymmetry from the decay of $n_{R_1}$ into any leptonic
flavour, though the most dominant contribution would come from the $\tau$
direction. Note that although we gain from a large Yukawa coupling
$\lambda_{33} \sim 1$, we pay the price of a very small $M_{n_{R_3}} \sim$ keV
in the loop. An order-of-magnitude estimate is
\bea 
\epsilon_1 \simeq {1\over 6 \pi}
 |\lambda_{33}|^2 \left(\frac{M_{n_{R_3}}}{ M_{n_{R_1}}}\right) 
\ln\left(\frac{M_{n_{R_3}}}{ M_{n_{R_1}}}\right) 
\delta,
\label{epsscn1}
\eea
where $\delta$ captures the CP violating phases which can be order one.  On
the other hand, when we have $n_{R_2}$ inside the loop, we lose in the
smallness of the Yukawa coupling ($\sim 10^{-5}$), but we gain in the loop
factor. Still, the contribution from $n_{R_3}$ exchange dominates over the one
from $n_{R_2}$ exchange\footnote{Since $n_{R_3}$ is in the keV range, 3-body
  decays like $n_{R_1} \!\to \!  \ell \bar{\ell}\!\ n_{R_3}$ are possible.
  However, this channel will give a null CP asymmetry.}.  It is worth noting
that the bound derived by Davidson and Ibarra \cite{di} on CP asymmetry is not
applicable because what we are considering is the decay of the
next-to-lightest RH neutrino, and not the lightest one.  Putting numbers, the
CP asymmetry can be $\epsilon_1 \sim 5\cdot 10^{-9}$, which is still quite
interesting.  This can be further enhanced by invoking e.g. resonance effects
\cite{pilaftsis}.

An important point is that although the Yukawa coupling $\lambda_{33}$ is
large, the light active mass generated through the mechanism of
Ref.~\cite{delaguila} is small (less than an eV), hence the washout effect
from $\Delta L = 2$ process via $n_{R_3}$-exchange graph would be of little
numerical significance. This turns the scenario to be very attractive.  
The relation between the baryonic asymmetry and the
 leptonic remains the usual one, given by  
\bea 
Y_{B} = - \left({8N+4\over
 14N+9}\right)\sum_{\alpha=e,\mu,\tau} Y_{L_\alpha}, 
\eea
where the number of generation is $N=3$.

\subsection{Scenario $\mathbf {II}$: A 4-generation case}
Our second scenario consists of invoking a fourth chiral family, i.e. each
entry of the mass matrix (\ref{m2}) is actually a (4$\times$4) matrix.  We
assume that in this picture all four RH neutrinos weigh in the TeV range. We
mention here that in spite of stringent constraints from LEP electroweak
measurements, there is still a window left for the fourth family (see
\cite{fourgen}).  We emphasize that in the context of leptogenesis the
r\^{o}le of the fourth lepton doublet is not much different from that of a
supersymmetric partner, so the leptogenesis consequences of a four generation
scenario are indicators of a more generic picture.

We now try to see what are the constraints on the different elements of the
Yukawa matrix. Recall, the first index of $\lambda$ corresponds to the RH
neutrino and the second one to the LH active neutrino.  The requirement that
light active neutrino masses corresponding to the first three families is less
than $\sim$ 0.05 eV constrains all the elements of the first three columns to
be less than $\beta \sim 10^{-5}$. If we assume that the generated CP
asymmetry is due to the decay of $n_{R_1}$, then all entries in the first row
should be less than $\epsilon \sim 10^{-7}$. The remaining elements,
$\lambda_{i4}$ for $i = 2,3,4$, have to satisfy the requirement that the
fourth generation active neutrino has to weigh above $M_Z/2$. To ensure this
one must have each $\lambda_{i4}$ at least $\alpha \sim 1.4 \times \pi$,
assuming the RH masses are around 500 GeV (Note, as mentioned already,
$\lambda_{14}$ receives a stronger constraint $< 10^{-7}$, necessary for
out-of-equilibrium decay of $n_{R_1}$).  Recall that perturbation theory is
assumed to be valid as long as $\lambda^2/4\pi$ remains approximately within
unity. Our requirement for $\lambda_{i4}$ for $i=2,3,4$ barely exceeds that
limit.  Actually the radiative origin of the LH active mass, even for the
fourth generation, is responsible for pushing the Yukawa couplings to such
large values.  It could have been avoided by adding an extra RH singlet
neutrino for the fourth family and switching on the $\Phi_2$ interaction in
Eq.~(\ref{lag-yuk}). However, for the moment we stick to our present scenario.
So the Yukawa matrix looks like (again, with all entries subject to order one
uncertainties)
 \bea
 {\cal{\lambda}}=\left(\begin{array}{c c c c} 
 \epsilon  & \epsilon & \epsilon & \epsilon\\
 \beta & \beta & \beta & \alpha \\
 \beta & \beta & \beta & \alpha \\
 \beta & \beta & \beta & \alpha 
\end{array}\right) . 
\label{lam4} 
\eea

Now we turn our attention to CP asymmetry, which we assume to be generated by
the decay of the lightest RH neutrino $\nu_{R_{1}}$.  We have seen that we
must require some large couplings to satisfy the fourth generation neutrino
mass constraints: $\lambda_{24} \sim \lambda_{34} \sim \lambda_{44} \sim
\alpha$. These large couplings enhance CP asymmetry, see Eq.~(\ref{eps1}).
{\em This is precisely the reason why a fourth family has been added}. We
assume that all RH masses weigh in the range 500 GeV to 1 TeV. This obtains
\bea 
\epsilon_1 \simeq {9\over 16 \pi}
|\beta| |\alpha| f \delta.
\label{epsscn14}
\eea
The loop function $f$ turns out to be order one for our choice of RH masses.
All the phases are contained in $\delta$ which can be pushed towards its
maximum value of unity.  So we gain both in the Yukawa coupling and in the
loop factor, contrary to what happens in scenario I.  Putting numbers, we may
expect $\epsilon_1$ to be a few times $10^{-6}$.  We have taken note of the
fact that the $\Delta L = 2$ processes involving the fourth family of leptons
in external legs are rapid and hence in thermal equilibrium.  It is
intuitively easy to see this by considering such a process with fourth
generation active LH neutrino ($\nu_4$) in external legs ($\nu_4 \phi \to
\nu_4 \overline{\phi}$ with $n_R$ exchange). This is equivalent to the see-saw
diagram that produces the heavy ($>$ 45 GeV) mass for $\nu_4$. Consequently,
the lepton asymmetry in the fourth leptonic direction is washed out. This
modifies the relationship between baryon and lepton asymmetry, which we obtain
as
\bea
Y_{B} = - \left({8N+4\over
 14N + 25}\right)\sum_{\alpha=e,\mu,\tau} Y_{L_\alpha},
\eea
where $Y_{L}$ is the produced leptonic asymmetry only for the light active
leptonic flavours, but $N = 4$ is the total number of generations.

\section{Observations and Outlook}
\begin{enumerate}
\item We have explored the {\em liason} between little Higgs mechanism and TeV
scale leptogenesis in two ways.  The first one relies on 3 generations only
but the mass textures for the third generation is different from the first
two.  The second approach is a generic extension of the particle content by
invoking a fourth chiral family, which opens the window for a more general
framework like supersymmetry.  A detailed description of flavour mixings among
the light active neutrinos is beyond the scope of the present work. However,
we emphasize that a hierarchical pattern of light active neutrino states is
necessary to enhance the CP asymmetry.  This can be arranged by adjusting the
order one uncertainties in different elements of the Yukawa coupling matrices
in Eqns.~(\ref{lam3}) and (\ref{lam4}), ensuring agreement with the
oscillation data and WMAP constraints.

\item Although $N_L$ is an SU(2) singlet, a lepton asymmetry cannot be
  generated from its decay. Its r\^{o}le is to influence the mass matrix in
  Eq.~(\ref{m2}) the way we have shown for individual generations.
  
\item Does $n_R$ decay into the SU(2) singlet $N_L$, thus opening a new
  channel for leptogenesis?  For this we look into the Yukawa interaction
  involving $N$, $n_R$ and the radial excitation $\rho_1$ (see
  Eqs.~(\ref{phi12}) and (\ref{su2-lag}))
\bea
 -L_{\rho} =  \bar n_{iR}\left( 1-{h^\dagger h\over 2 f^2}\right) 
  \rho_1^*\lambda_{ij}N_{jL}\ + 
 {\rm h.c.}\ .
 \label{rad} 
 \eea 
 But the decay $n_R \to N \rho_1$ is kinematically either disallowed or
 suppressed since $M_{\rho_1} \sim $ TeV.

\item Light active neutrino data provides an access to the Majorana mass of
  $N_L$'s. In scenario II this mass turns out to be in the keV range for the
  first three generations, while the fourth $N_L$ would weigh around 100 TeV
  \footnote{Such a high scale can be accomodated by extending the simplest
    little Higgs framework, see \cite{Schmaltz3}.}.  Its decay is however in
  equilibrium and will not produce any leptonic asymmetry.
 
  The $N_L$'s can mix with $\nu_L$'s but this mixing ($\sin \theta_m \sim
  v/f$) can be kept just consistent with the few per mille precise LEP data on
  $Z$ invisible decay width \cite{lep}.

\item As has been shown in the context of 331 models \cite{foot,frampton} and
  for the simplest little Higgs model \cite{kong}, given the assigned charges
  for the fermion fields each generation is anomalous. However, an anomaly
  free model is obtained if the number of generations is a multiple of 3.  In
  general, one may argue that the UV completion will make the model
  anomaly-free \cite{Schmaltz1}. In this case the four generation model
  proceeds as discussed above.  Alternatively one could add fermionic
  particles to each generation to ensure it is anomaly free per generation as
  was done in \cite{sanchez}.

\item Besides the light active physical state $\nu' \simeq \nu + (v/f) N$, we
  also have the orthogonal, dominantly sterile, $N' \simeq N - (v/f) \nu$
  state which couples to the gauge bosons. These would lead to additional
  missing energy signatures at colliders.

\item In the context of the littlest Higgs model \cite{cohen}, a recent paper
  discussed the production of the neutrino masses \cite{Hanetal} without
  having RH neutrinos.  We note that if RH neutrinos are included, then it may
  be possible to produce a lepton asymmetry with only 3 generations consistent
  with observed data as the smallness of the neutrino masses arising via the
  interaction with the triplet field can be due to a small vev for the
  triplet. This allows some of the Yukawa couplings entering the expression of
  the CP asymmetry to be unsuppressed \cite{hambye3,Hambye} compared to the
  usual analysis of TeV scale models of leptogenesis with 3 generations. The
  details of this scenario will be discussed elsewhere.

\end{enumerate}

{\em To conclude}, the treasures of the TeV scale could include both the
little Higgs scenario and thermal leptogenesis.  In this paper, we have
studied the synergy between these two mechanisms. The little Higgs scenario
has direct consequences on the structure of neutrino mass matrix. The
radiative origin of the light active neutrino masses is an interesting feature
of our scenario. We have used the constraints derived from there to embed
leptogenesis in the simplest little Higgs framework.

\section*{Acknowledgments} 
We thank S. Vempati, B. Gavela, and T. Hambye for interesting discussions and
S. Davidson and A. Pilaftsis for making valuable comments on the manuscript.
GB acknowledges hospitality of LPT-Orsay and SPhT-Saclay, while AA and GB
thank the CERN Theory Division for hospitality where the work was initiated.
GB's research has been supported, in part, by the DST, India, project number
SP/S2/K-10/2001.

\end{document}